\documentclass[twocolumn]{aastex62}

\def\simgt{\lower.5ex\hbox{$\; \buildrel > \over \sim \;$}}
\def\simlt{\lower.5ex\hbox{$\; \buildrel < \over \sim \;$}}
\newcommand\tbce{T$_{\rm bce}$}

\newcommand{\msun}{\ensuremath{\, {M}_\odot}}
\newcommand{\Msun}{\ensuremath{\,{M}_\odot}}
\newcommand{\Lsun}{\ensuremath{\,{L}_\odot}}

\newcommand{\Yin}{Y$_{\rm in}$}
\def\he3{$^3$He}
 
\graphicspath{{./}{figures/}}

\received{XXXX}
\revised{XXX}
\accepted{XXXX}
\submitjournal{ApJ}

\shorttitle{Lithium in NGC\,2808}
\shortauthors{D'Antona et al.}

\begin{document}

\title{\bf The Lithium test for multiple populations in Globular Clusters:\\ Lithium in NGC\,2808}
\correspondingauthor{Francesca D'Antona}
\email{francesca.dantona@inaf.it; franca.dantona@gmail.com}

\author{Francesca D'Antona}
\author{ Paolo Ventura} 
\affiliation{INAF-- Osservatorio Astronomico di Roma, Via Frascati 33, 00078 Monteporzio Catone (Roma), Italy}
\author{ Anna Fabiola Marino} 
\affiliation{Research School of Astronomy \& Astrophysics, 
ANU, Mt Stromlo Observatory, Cotter Rd, Weston, ACT 2611, Australia}
\affiliation{ Dipartimento di Fisica e Astronomia ``Galileo Galilei'', Univ. di Padova, Vicolo dell'Osservatorio 3, Padova, IT-35122}
\author{Antonino P. Milone}
\affiliation{ Dipartimento di Fisica e Astronomia ``Galileo Galilei'', Univ. di Padova, Vicolo dell'Osservatorio 3, Padova, IT-35122}
\author{Marco Tailo} 
\affiliation{ Dipartimento di Fisica e Astronomia ``Galileo Galilei'', Univ. di Padova, Vicolo dell'Osservatorio 3, Padova, IT-35122}
\author{Marcella Di Criscienzo}
\affiliation{INAF-- Osservatorio Astronomico di Roma, Via Frascati 33, 00078 Monteporzio Catone (Roma), Italy}
\author{Enrico Vesperini}
\affiliation{Department of Astronomy, Indiana University, Bloomington, IN 47401, USA}


\begin{abstract}
In the globular cluster NGC2808, a quasi-standard initial lithium abundance is derived for a red giant belonging to the `extreme' population, characterized by a large helium overabundance, and by abundances of proton--capture elements typical of nuclear processing in gas at very high temperatures, where the initial lithium has been fully destroyed. 
The observations of lithium in such extreme cluster stars are important to test different models for the formation of multiple populations  in old Globular Clusters. 
In the asymptotic giant branch (AGB) scenario, fresh lithium is synthetized during the initial phases of hot bottom burning which, afterwards, synthetize the other p-capture elements. 
We model the abundance of lithium in the ejecta of super--AGB models, finding values consistent or larger than observed in the `extreme' giant; these same models describe correctly the magnesium depletion and silicon enrichment of the extreme population of NGC\,2808, so the overall agreement provides further support to the AGB scenario.
In the models involving massive or supermassive stars, the Lithium observed requires a mixture of the lithium--free ejecta of the polluting population with more than 40\% of standard-lithium pristine gas. 
The extended chemical anomalies of NGC\,2808 stars are then to be all explained within at most 60\% of the possible dilution range, the initial helium mass fraction in the ejecta should be Y$\simgt$0.5, to account for the Y$_e\sim$\,0.38--0.40 of the extreme population, and further observations of p--process elements are needed to check the model.
\end{abstract}


\section{Introduction} 
\label{SEC_intro}
Two main environments can be the site of the nuclear proton-capture reactions necessary to explain the chemical patterns \citep[e.g.][]{carretta2009a} of multiple populations in globular clusters (GCs): the gas of the `second generation' (2G) stars may have been processed 

{\it 1)} either in the H--burning convective cores of `first generation' (1G) high mass \citep[e.g.][]{decressin2007, demink2009, bastian2013} or supermassive stars \citep{denissenkov2014} ---hereinafter the Convective Core Hydrogen Processing (CCHP) models.

{\it 2)} or  in the  ``Hot Bottom Burning" (HBB) high temperature layers at the base of the convective envelopes of 1G massive Asymptotic Giant Branch (AGB) stars \citep{ventura2001} and ``super--AGB'' stars \citep{siess2010, ventura2011sagb}  ---hereinafter the AGB scenario. 

In this work we reconsider two important signatures of 2G stars, for which the CCHP models and the AGB scenario make different predictions:

{\it i)} The observed color magnitude diagrams of GCs allow to put  a strict upper limit  \citep[\Yin$\sim$0.41, according to][]{chantereau2016} to the initial helium mass fraction in the gas forming 2G stars, both from the modellization of the Horizontal Branches and from the observations of the Main Sequence (MS) and Turnoff morphologies.
As for the CCHP models, it is important to point out that the phase of core-H-burning generally goes throughout full H-exhaustion leading to extreme helium abundances significantly larger than those observed. The presence of an upper limit on the helium content thus  requires very specific and ad hoc assumptions concerning an early end of the H--burning when the helium core abundance goes beyond such a value \citep[e.g.][]{gieles2018}. On the contrary, the helium content of massive AGB envelopes is directly linked to the second--dredge up (2DU) phase, is limited to Y=0.35--0.38 in the standard models \citep[e.g.][]{ventura2010, doherty2014}, and reaches up to Y$\sim$0.40 in rotating models \citep{georgy2013,choi2016}.

{\it ii)} CCHP models fully destroy lithium. When this fragile element is found in the atmospheres of 2G stars, the gas  processed by p-capture in CCHP models must be heavily diluted with unprocessed gas still preserving its standard population II  abundance. This explanation is indeed possible for a number of cases \citep[e.g.][]{dantona2012li}, but it must be carefully examined for extreme 2G stars. 
In the AGB scenario, at the beginning of the HBB phase, fresh lithium is produced by the \cite{cameronfowler1971} mechanism, and remains at very high abundances in the envelope until  $^3$He is totally consumed, and eventually it is fully burned. A quantitative and close comparison with the data to either support or rule out the model, is in general complicated, since the lithium average abundance in the ejecta is scarcely constrained, due to its strong dependence on the mass loss rate during the lithium rich phase \citep{ventura2005a}. 

In Sect.\,\ref{2808}  we focus our attention on a red giant \citep[identified with the number  \#46518 in the catalogue by][]{dorazi2015} characterized both by the maximum helium and by the presence of lithium. 
 \cite{dorazi2015} find several lithium rich giants, but  \#46518 belongs to the `extreme' population ---the group whose helium abundance is at the highest values--- 
and its lithium abundance is a bare factor two smaller than in the 1G giants of the same sample. 

In Sect.\,\ref{AGBmodels} we compute and analyse the detailed lithium nucleosynthesis in our super--AGB models whose yields match the other chemical anomalies in the extreme stars of NGC\,2808  \citep{dicrisci2018}, and confirm that the models provide the lithium and p--processing required. In Sect.\,\ref{dilution}  we consider the conditions under which this giant can be explained by the CCHP models, and find tight constraints and very specific dilution patterns are required. In the Sect.\,\ref{conclusions} we summarise the results and conclude that, with the help of future observations, the `lithium test' may result to be the ultimate tool to discriminate among formation models for GC stars.

\begin{figure}
\vskip -0.8cm
\hskip -20pt
\includegraphics[width=1.1\columnwidth]{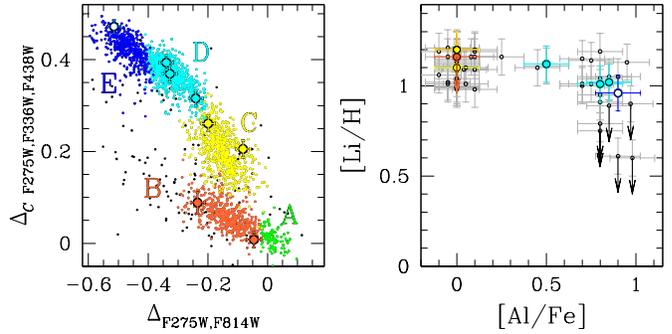}
\vskip -6.5cm
\caption{\textit{Left: chromosome map ---location of stars in the pseudo--colors combination deviced by \cite{milone2017chromo}--- for the giant stars in NGC\,2808, subdivided into 5 groups from A to E by \cite{milone2015},  highlighted in different colors. 
Open black circles  with error bars are the giants  in NGC\,2808 having both lithium determination and the HST photometry to locate them on the map. The giant \#46518 is shown by the black circle with white filling at the upper edge  of the  chromosome map, well inside the E group. Right: in the lithium versus aluminum plane the stars of the left panel are shown in color, along with the others from the sample of \cite{dorazi2015}. }}
\label{f1}
\end{figure}

\section{The case of NGC 2808}
\label{2808}
The abundance patterns of the GC NGC\,2808 constitute a valuable benchmark for studying multiple populations.  In NGC\,2808, all the high temperature p-capture elements show large abundance variations, and the stars with different patterns of chemical abundances are subdivided into 6 discrete groups \citep{carretta2015, carretta2018}. Also the description of this cluster in terms of `chromosome maps' \citep{milone2017chromo} highlights the presence of at least 5 different discrete populations \citep[named with letters from A to E in][see Figure\,\ref{f1}]{milone2015}. 
     In spite of this complexity, this cluster is indeed the best prototype of a `simple' evolution leading to the formation of multiple populations, as its stars do not show clear signs of iron spread. 
     In the recent reanalysis of high dispersion spectroscopic data for  NGC\,2808, \cite{carretta2018} point out that the detailed correlations between the different elements do not allow a simple explanation of abundances involving a single pollutor source.

\subsection{NGC 2808 in the AGB scenario}
NGC\,2808 displays a triple MS \citep{dantona2005, piotto2007}  well explained by assuming three stellar populations with different helium abundances. In fact both the  ``blue" main sequence (MS) and a number of ``blue hook" stars  at the hot end of the Horizontal Branch (HB) \citep{dcruz1996} indicate that NGC\,2808 contains $\sim$10\% of stars with a helium mass fraction Y$\simeq$0.35--0.40 \citep{dantona2005}, values tantalizingly close to the maximum helium in the ejecta of AGB models. X--shooter spectroscopy of two MS stars belonging to the standard and to the blue MS \citep{bragaglia2010} confirm that the blue MS star composition is typical of an extreme ---highly p-capture processed--- population.

The triple MS found a natural explanation in the work \cite{dercole2008} exploring the formation of the 2G in the AGB scenario: 1) the red MS is the standard helium MS of the 1G; 2) the stars born from `pure' massive AGB ejecta, preserve the maximum chemical anomalies and the maximum possible helium content form the blue MS; 3) the `intermediate' MS contains stars born by gas mixture of AGB ejecta and pristine gas re-accreted from the interstellar medium. Further analysis of the chemical composition of stars in this cluster by a chemical evolution model \citep{dercole2010, dercole2012} were able to integrate the first proposal. The chromosome map ---see Figure \ref{f1} and \citet{milone2015}--- points to a more complex but similar scheme of formation of the multiple generations \citep{dantona2016} in which the extreme stars (group E, blue in Fig.\,\ref{f1})  are born in a cooling flow collecting the {\it undiluted} envelopes lost by the heaviest AGB and super--AGB stars: the composition of these stars in the AGB scenario simply reflects the nucleosynthesis products of the 2DU (for the helium content) and of HBB. 
\begin{figure}
\vskip-1cm
\centering
\includegraphics[width=0.99\columnwidth]{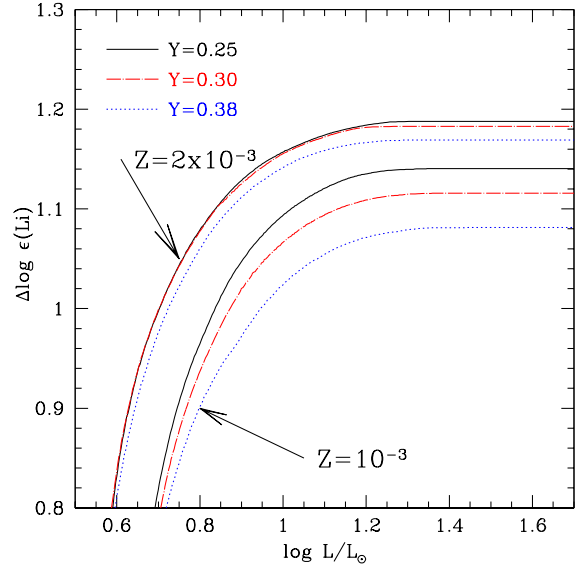}
\vskip -2.cm
\caption{\textit{$\Delta \log \epsilon$(Li) is the logarithmic difference between the initial abundance of lithium and the abundance along the red giant branch evolution of the masses (Table 2) evolving at 12\,Gyr for different helium and metallicity. The value reaches a maximum when the first Dredge Up, at the base of the giant branch, is complete. }}
\label{f2}
\end{figure}
\begin{table}
\centering
\caption{ Standard dilution of lithium at first Dredge up at age 12Gyr. The evolving mass and the difference in logarithmic abundance with respect to the initial content are given for three helium values and two metallicities. }
\begin{tabular}{c c c c c }   
\hline
& \multicolumn{2}{c}{Z=10$^{-3}$} & \multicolumn{2}{c}{Z=2 $\times 10^{-3}$} \\
$Y$  &  $M/M_\odot$  & $\Delta \log \epsilon(Li)$  &  $M/M_\odot$  & $\Delta \log \epsilon(Li)$    \\
\hline
0.25 & 0.814  & 1.14 & 0.834 &  1.19   \\
0.30 & 0.748 &  1.12 & 0.762 &   1.18  \\
0.38 &  0.644 & 1.08 & 0.658 & 1.17   \\      
\hline     
\end{tabular}
\label{tabyield}
\end{table}

\subsection{NGC\,2808 in CCHP models}
The CCHP models have some difficulty in describing that 2G formation in NGC\,2808 occurs in discrete events \citep{renzini2015}. In addition, \cite{carretta2018} show that the abundances of p-capture elements are not consistent with a single dilution scheme.
Anyway, its extreme population is not necessarily made up from pure ejecta \citep[see, e.g.][]{denissenkov2014}. Uniform dilution with an $\alpha$\ fraction of pristine gas may occur before second generation formation begins. In this case, it is sufficient to assume that the core-hydrogen burning stops at a helium content Y$_{max}$\ larger than the Y$_e \sim$0.4 of the extreme population. 
Assuming Y=0.25 for the pristine gas, we get:
\begin{equation}
Y_{max}={{ Y_e-0.25\alpha} \over {(1- \alpha)}}
\end{equation}
Notice that the initial dilution is a further hypothesis to be added in the models, as  the diluting gas must in all cases contribute in the right amount to leave an Y$_e \sim$0.38--0.4, similar for all the clusters hosting an extreme generation.

\subsection{Lithium in the giants of NGC 2808}
\label{2808li}
 \cite{dorazi2015} examined the lithium and aluminum abundance in a large group of  NGC\,2808 giants, less luminous than the red giant bump. The sample choice guarantees that  the lithium content in atmosphere has only suffered standard convective dilution at the first dredge up \citep{iben1964firstdu}, as the stars are subject to further depletion due to additional mixing mechanisms \citep[e.g.][]{charbonnelzahn2007}  above the red giant bump. 
The left panel of Fig.\,\ref{f1} shows a few of the giants for which measurements of lithium are available on the NGC 2808 chromosome map (Marino et al. 2018, submitted, see also \cite{carretta2018}). Four stars belong  either to the 1G population (group B, dark orange)  or to group C (yellow), which formed from matter very highly diluted with pristine gas. Their  [Al/Fe] is low, confirming they belong ---or have abundances similar--- to 1G, and their lithium abundance is $\log \epsilon$(Li)$\sim$1--1.2\footnote{We use the usual notation $\log \epsilon(Li)=\log(N_{Li}/N_H)+12$, that is the number abundance with respect to hydrogen, posing the H number abundance at 10$^{12}$.} (Fig.\,\ref{f1}, right panel).
Three other stars in \cite{dorazi2015} sample belong to the group D (cyan in  Fig.\,\ref{f1}), a population in which the p--processed ejecta suffered intermediate dilution with pristine gas, and 
the giant ID\#46518, belongs to the `extreme' group (E), and has $\log \epsilon$(Li)=$0.96 \pm 0.1$. 
The giants with strong aluminum display either abundances slightly smaller that the 1G values ($\log \epsilon$(Li)$\simeq$1) or much lower upper limits. 
We remark that the aluminum content does not allow an easy discrimination between groups D ---corresponding to group I2 in \cite{carretta2018}--- and E. In fact, although the average abundance of aluminium in the E group is given as [Al/Fe]=1.292$\pm$0.029 (their Table 3), the star-to-star abundances are very scattered (see their Fig.\,3) and partially superimposed to the abundances of group I2. On the other hand,  the chromosome map in Fig.\,\ref{f1} shows unambiguously that the  giant \#46518 belongs to the extreme population. 

We estimated the convective dilution at 1st dredge up in coeval models of giants with different helium content, for two different metallicities bracketing the value appropriate for NGC\,2808. Figure \ref{f2} and Table 1 show that the lithium depletion  depends slightly on Y and Z, at a level $<$0.1\,dex. 
The corrections in Table 1 are applied to the giant ID\#46518: as it belongs to the high--helium E group, we use the Y=0.38 models. We obtain for the initial lithium of the star log\,$\epsilon$(Li)= 2.04 with $\sim$25\% total uncertainty. 
By applying the correction for Y=0.25, the initial abundances of the B and C population stars (standard Y) is log\,$\epsilon$(Li)= $2.2 \div 2.4$, values well compatible with the abundance measured at the surface of population II dwarfs  \citep{asplund2006}.

\section{Super--AGB model results}
\label{AGBmodels}
\subsection{Lithium production in AGB and super--AGB models }
The problem of lithium production in the envelopes of luminous AGB stars has been examined more than 40 years ago, based on the chain proposed  by \cite{cameronfowler1971}. 
In fact, the terminology ``hot bottom burning'' (HBB) was first used for the envelope models of AGBs \citep{scalo1975}, in which the temperature at the bottom of the convective envelope (\tbce) reached \tbce$\sim$40\,MK, activating the chain $^3$He($\alpha,\gamma)^7$Be. \cite{ventura2001} recognized that \tbce\ could reach values much larger than 40\,MK, at least in the massive low metallicity AGBs. This allows full CNO cycling and the other p-capture reactions which characterise the composition of second generation stars in GCs \citep[see, e.g., the discussion in ][]{dantona2016}. Lithium yields from HBB for GC type metallicities were provided by \cite{ventura2008aa}.  

\begin{figure*}
\vskip -2.5cm 
\centering{
\includegraphics[width=15cm]{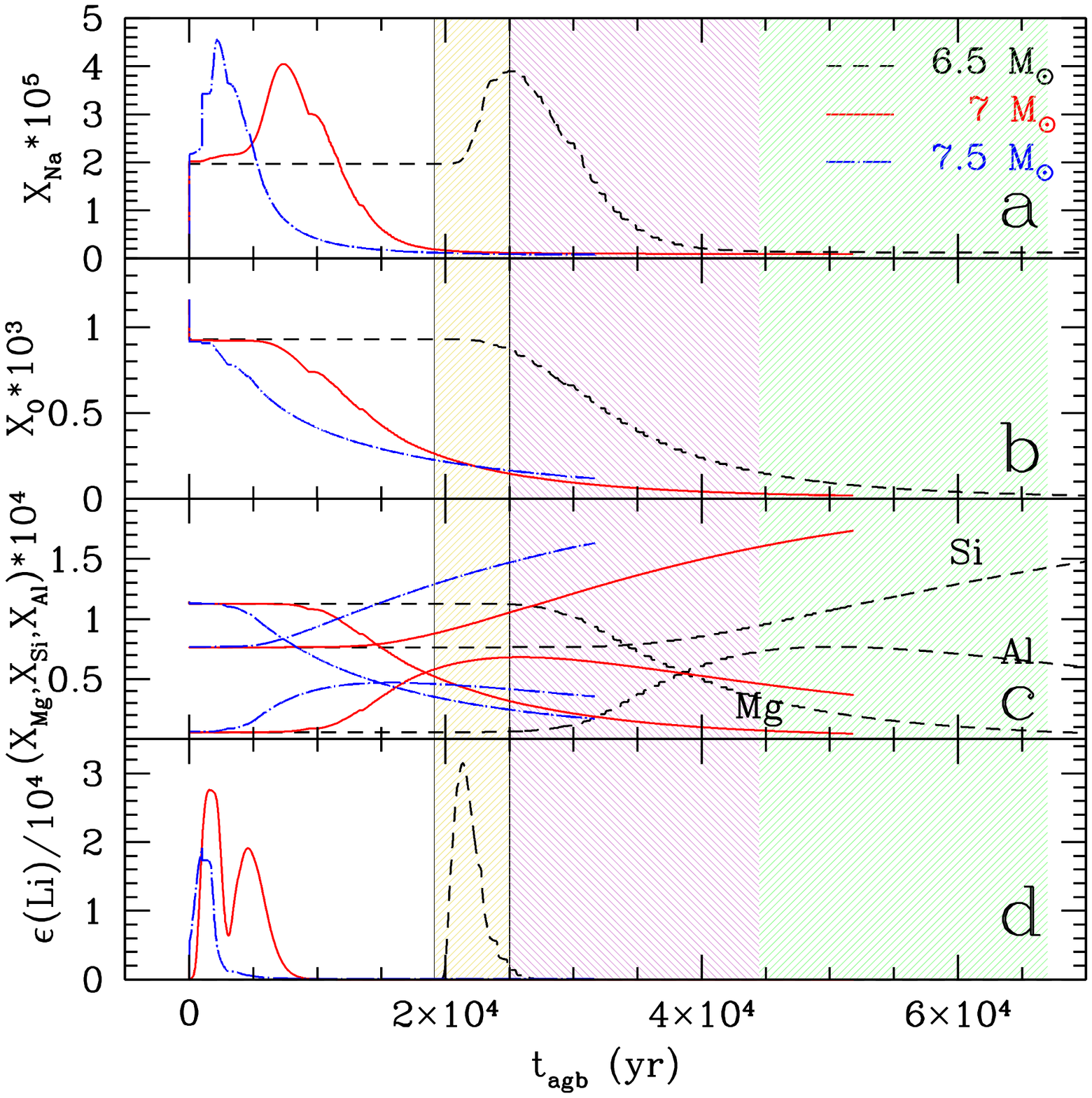}
}
\vskip -4.6cm 
  \caption{\textit{Evolution of the abundances of sodium (panel a), oxygen (b), magnesium, aluminum and silicon (c), and lithium (d), at the surface of super--AGB stars of 6.5\Msun\ (dashed), 7 (full line) and 7.5\Msun\ (dash--dotted). For the 6.5\msun\ we highlight in different colors the time of production and destruction of lithium (gold), the main epoch of  sodium, oxygen and magnesium depletion, plus and aluminum production (purple) and the longer time of final increase of silicon at the expenses of magnesium and aluminum (green). 
}}
  \label{f4}
\end{figure*}

In their first super--AGB models, \cite{ventura2011sagb} found a huge lithium abundance in the ejecta (for 6.5$\leq$M/\msun$\leq$8, see their Table 2). This was a consequence of the \cite{blocker1995} mass loss rate,  calibrated on the population synthesis of the L\,$\sim 2\cdot 10^4$\Lsun\ sample of lithium rich AGBs in  the Magellanic Clouds  \citep{ventura2000} and applied to super--AGBs at L\,$\sim 10^5$\Lsun. As the dependence on the luminosity is very strong ($\propto L^{3.7}$), the huge mass loss rate did not allow time for the operation of the ON cycle, and for the p-captures on Magnesium, so the resulting nucleosynthesis was not compatible with that required for the extreme second generation stars.

\begin{table}
\centering
\caption{ Average abundances in the Super--AGB ejecta for different masses and mass loss rates }
\begin{tabular}{c c c c c c c c c}   
\hline
 \footnotesize{M}  &  $\eta$  & $\log \epsilon(Li)$  &  \footnotesize{[O/Fe]} &  \footnotesize{[Na/Fe]} &  \footnotesize{[Mg/Fe]}  &  \footnotesize{[Al/Fe]} &  \footnotesize{[Si/Fe]} \\
\hline
\hline
6.5 & .02  & 2.20 & --0.38 & 0.37 & --0.09 & 1.18  & 0.44  \\
6.5 & .01  & 2.19 & --0.56 & 0.22 & -0.35  & 1.05  & 0.48  \\
6.5 & .03  & 2.42 & --0.38 & 0.49 & --0.15 & 1.19   & 0.43 \\
7.0  & .02 & 2.50 & --0.33 & 0.44 & --0.04 & 1.07   & 0.48 \\
7.5 & .02  & 2.82 & --0.21 & 0.48 &  0.02 & 1.02  & 0.42 \\
7.5 & .01 & 2.59 & --0.26 & 0.41 &  0.10  &  1.08 & 0.44 \\
7.5 & .03 & 2.51 & --0.36 & 0.51 &  0.02  &  1.26 & 0.43 \\
\hline     
\hline     
\end{tabular}
$^1$Initial abundances are [O/Fe]=0.4, [Mg/Fe]=0.4, [Si/Fe]=0.25,
[Na/Fe]=0, [Al/Fe]=0 
\label{tabyield}
\end{table}

\subsection{Lithium production in new super--AGB models}

Dealing with the problem of finding new ways to calibrate the mass loss rate in super--AGBs, \cite{dicrisci2018}  adopted a modulation of mass loss based on the different quantities of dust production for different parts of the evolution. This parametric approach was sufficient to achieve the required advanced nucleosynthesis shown by the extreme stars in NGC\,2808. Here we extend those computations to the lithium nucleosynthesis, for all the details see \cite{dicrisci2018}.

\begin{figure}
\vskip -1cm 
\centering{
\includegraphics[width=8cm]{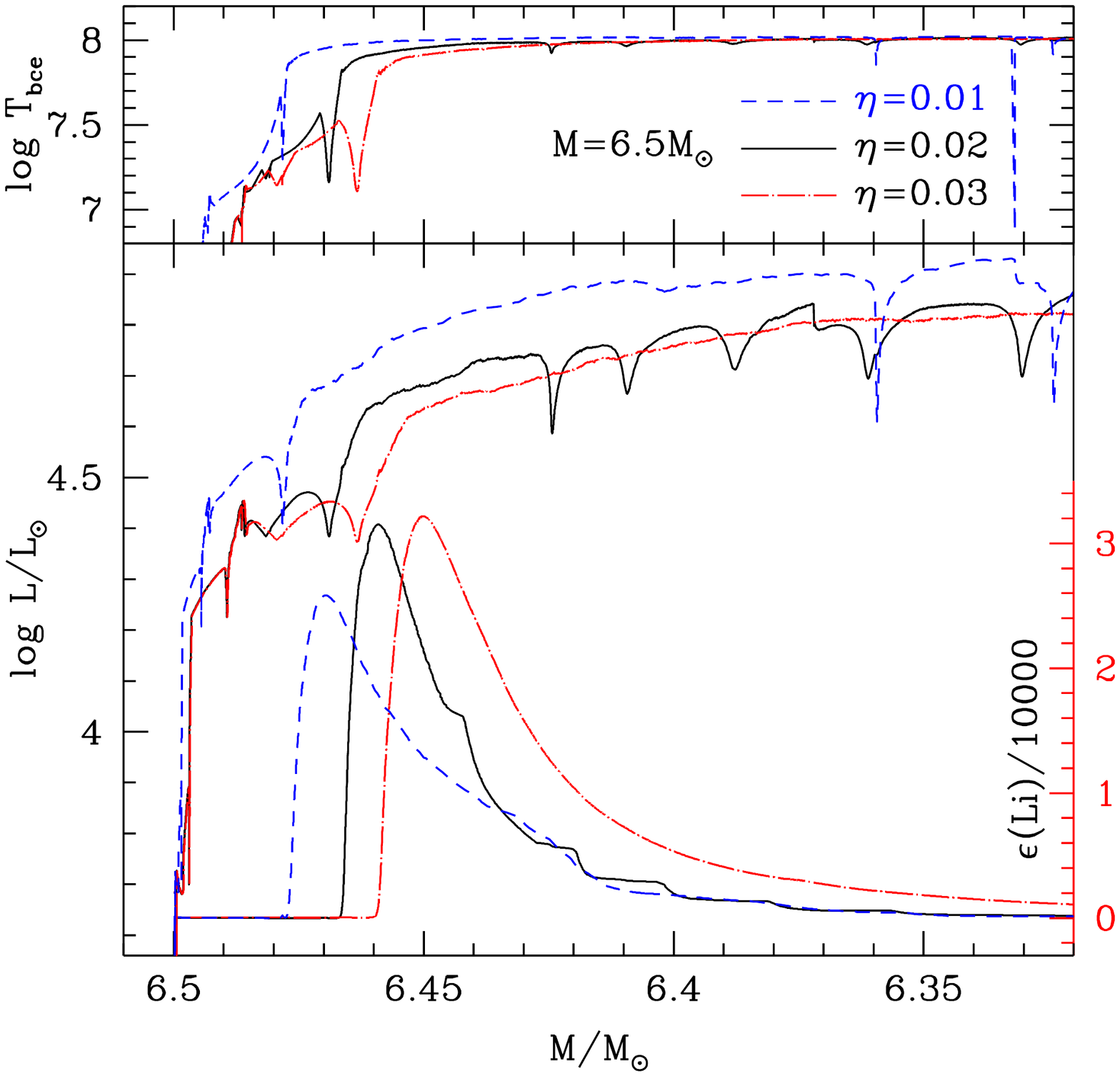}
}
\vskip -2.0cm 
  \caption{\textit{Evolution of the 6.5\msun\ tracks with different mass loss rates, $\eta$=0.02 (standard models, full black line), $\eta$=0.01 (blue dashed) and $\eta$=0.03 (red dash-dotted). Luminosity, lithium (bottom panel, scales on the left and on the right) and \tbce  (upper panel) are plotted versus total stellar mass (decreasing due to mass loss), covering the phase of lithium nucleosynthesis following the 2DU (recognized by the minimum in \tbce, followed by the onset of HBB).
    } }
 \label{f5}
\end{figure}

\begin{figure}[t]
\vskip -1cm 
\centering{
\includegraphics[width=8.1cm]{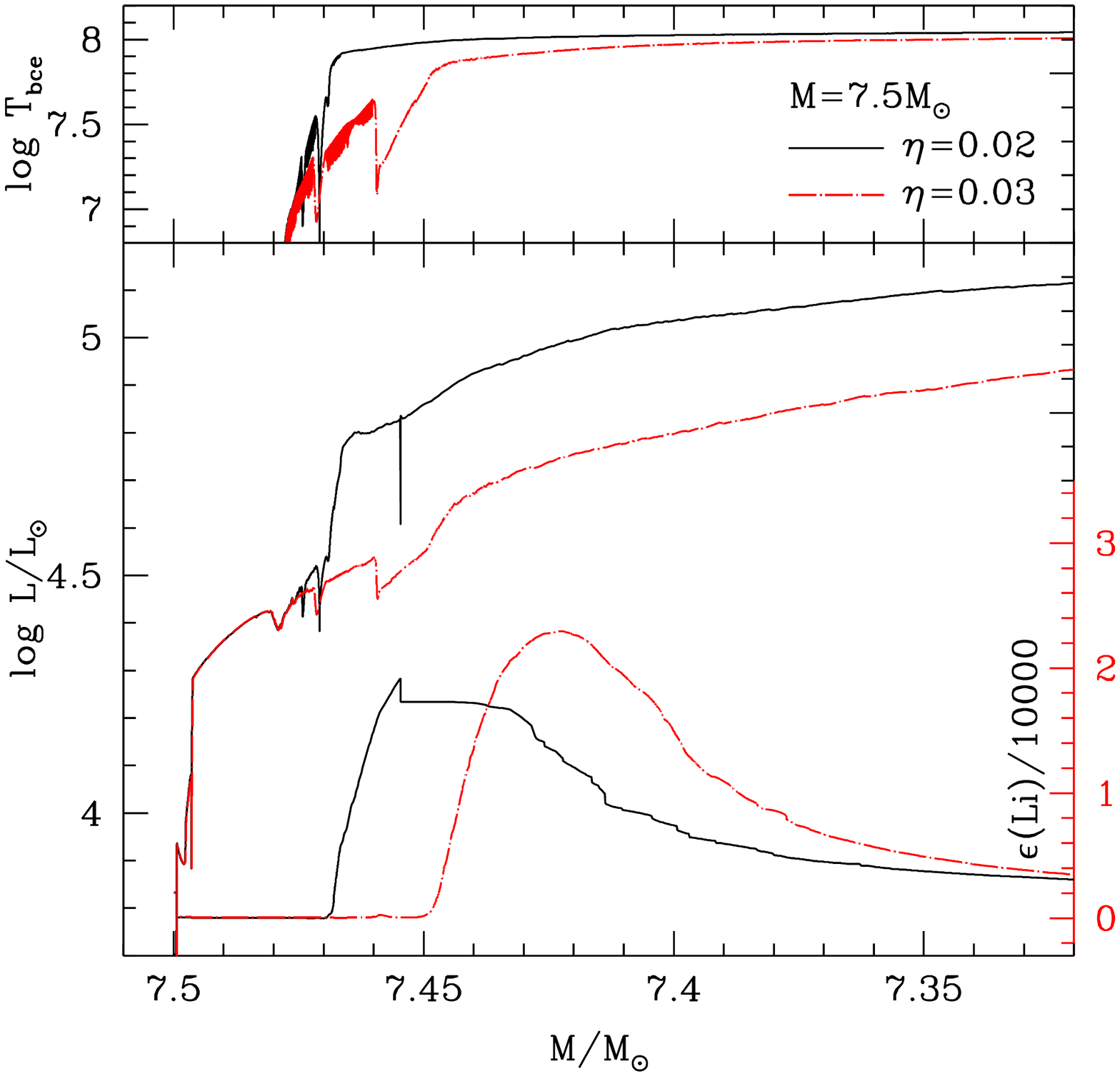}
}
\vskip -2.0cm 
  \caption{\textit{The same as Figure \ref{f5}, for the 7.5\msun\ track with  $\eta$ 0.02 -- black full line, and 0.03 --red dash--dotted.  The maximum lithium abundance reached in the 7.5\msun is smaller than in the 6.5\msun\ (Fig.\,\ref{f5}), but the total lithium ejected is larger, because the mass loss rate is much larger in the 7.5\msun.  
    } }
  \label{f6}
\end{figure}

The results are summarized in Table\,\ref{tabyield}, where $\eta$\ is the Reimer's parameter entering in the \cite{blocker1995} mass loss rate formulation.
The temporal evolution of different elements is shown in Figure \ref{f4} for masses 6.5, 7 and 7.5\msun, and highlighted for the 6.5\msun\ as example. In stars having such massive cores ($\sim 1.05$\msun) the HBB temperature increases fast after the 2DU has ended, and the lithium production stage (highlighted in gold) occurs previous to the beginning of the thermal pulses (this can be seen in Fig.\,\ref{f5} and \ref{f6}). The lithium phase occurs well before the main phase of depletion of  $^{16}$O,  $^{23}$Na and total Mg (region in purple). The phase of  $^{28}$Si production is the longest one, because it is also due to p-captures on aluminum, as shown in detail in panel c. The sodium peak  (due to the p-captures on the $^{22}$Ne dredged up) is concomitant to lithium production, while the phase of sodium depletion follows on the longer timescale of the other depletions. Thus the whole lithium production stage occurs during the first phases of evolution, and the following choices of mass loss affect the processing of the other elements (magnesium, oxygen, sodium) and not of lithium itself.

The larger masses 7 and 7.5 have similar behaviour, limited to a shorter total lifetime. The larger is the mass, the lower is the peak of lithium production. The reason is the faster cycle of production-desctruction of lithium due to the larger \tbce\ in the larger masses. In spite of the peak, the  lithium abundance in the ejecta is larger as the mass lost during the phase of lithium production dominates over the peak abundance reached. 

\subsection{Uncertainty in the lithium ejecta: changing the mass loss rate}
To understand the possible variations of lithium in the ejecta, we need a global study that goes well beyond the aim of this work. We exemplify the complexity of the problem by exploring different values of $\eta$.
Everything else being the same, we would expect that the lithium abundance depends linearly on the mass loss rate (that is on the $\eta$\ value). In fact the situation is more complex:  the $\eta$=0.02 and $\eta$=0.01 tracks for 6.5\msun\ do not show a simple behaviour (Fig.\,\ref{f5}). The track with smaller mass loss rate evolves to higher luminosity, and the total lithium lost is very similar. On the contrary, the average abundance increases by 0.2\,dex by increasing $\eta$\ by 50\%. In the largest super--AGB masses (the 7.5\msun, see Figure \ref{f6}),  the initially larger mass loss rate influences  \tbce\ and delays the lithium production, but the models are both cooler and less luminous, so, in the end, the total mass lost during the lithium rich phase is smaller.

Figures \ref{f5} and \ref{f6} give a fair illustration of how many physical inputs and parameters influence the lithium production, so that these results must be taken as guidelines more than at face values.
Notice that not all extreme stars may preserve the lithium of the ejecta, as additional mixing mechanisms, such as advocated by \cite{dicrisci2018} to explain the large spread in oxygen among the extreme stars in NGC\,2808, might begin to be activated even below the bump for very helium rich giants \citep[see][]{dantonaventura2007}. 

In conclusion, we should keep in mind that the results depend too much on the details of modelling and model inputs, and we should apply a generous uncertainty to the final average abundance. Still, the models show that a result very close to what we need for the extreme giant is obtained, within a few hundreths of dex.  

\section{Requirements of the CCHP models: dilution and lithium in 2G stars}
\label{dilution}
\begin{figure}
\vskip-1.5cm
\centering
\includegraphics[width=0.99\columnwidth]{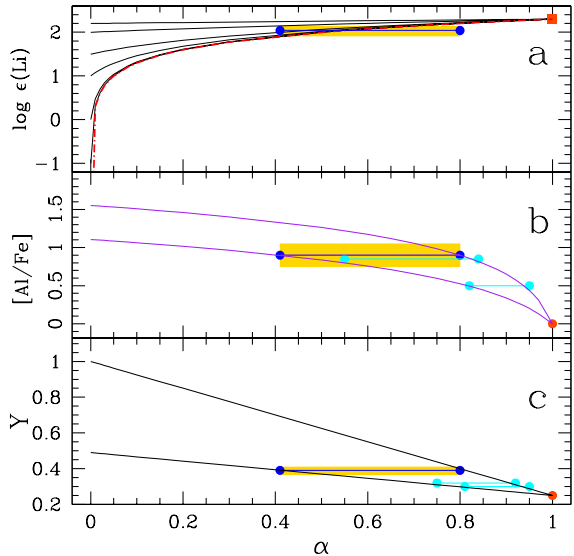}
\vskip -2.3cm
\caption{\textit{The abscissa is the fraction $\alpha$\ of gas with  standard population II abundances, mixed with a fraction (1--$\alpha$) of ejecta in the classic dilution scheme. In the three panels, abundances of lithium, aluminum and helium in the mixtures are plotted. For panel a (Lithium) the dilution curves correspond to values at $\alpha$=0 from log\,$\epsilon$(Li)=--1 to 2.2. The red dash-dotted line corresponds to fraction ($1-\alpha$) having $\epsilon$(Li)=0.  The point at  $\alpha$=1 has  log\,$\epsilon$(Li)=2.3. The golden areas cover the $\alpha$\ range allowed by the dilution model for the giant \#46518,  $0.4 \lesssim \alpha \lesssim 0.8$\ for lithium. Blue dots represent the extreme of the range for this giant.
The cyan and red dots show the range of initial lithium in the giants of group D and B--C, located at typical $\alpha$'s for their population. Aluminum (purple, scale on the left) and (probable) helium content Y  (black, scale on the right) are plotted, together with the dilution curves passing through $\alpha=0.4$\ and the values [Al/Fe]=0 and Y=0.25 at $\alpha$=1. 
}}
\label{f3}
\end{figure}

Figure \ref{f3} shows specific dilution curves for lithium (panel a), aluminum (b) and helium content (c) in mixtures of  a fraction $\alpha$\ (abscissa) of standard unprocessed gas with a fraction (1--$\alpha$) of ejecta. Thus
at $\alpha$=0 we read the abundances of undiluted ejecta from the polluters and at $\alpha$=1 we have the standard abundance  of the 1G population.
For lithium, the abundances chosen for the polluter gas are log\,$\epsilon$(Li)=--1, 0, 1, 1.5, 2, 2.2, and the 1G has the  log\,$\epsilon$(Li)=2.3 of population II dwarfs.  If the processed gas is not from AGB ejecta, the abundance at $\alpha$=0  will be $\epsilon$(Li)=0, and the red dot-dashed line represents lithium in the mixtures. 
The yellow region shows the  range of $\alpha$\ allowing to get the abundance of lithium in the giant \#46518 in this dilution model. 
The request of a minimum log\,$\epsilon$(Li)$\sim$1.9 gives a minumum $\alpha \sim 0.4$. Requiring that helium in this extreme giant is $\sim$0.4, implies a maximum possible dilution of $\alpha \sim 0.8$. 
The dilution curve for helium (panel c) are then defined by requiring an average Y=0.39 for the giant, at $0.4 \lesssim \alpha \lesssim 0.8$, and Y=0.25 for the 1G giants at $\alpha$=0. The plot shows the limiting dilution lines passing through $\alpha=0.4$\ and $\alpha=0.8$.
The dilution curve for aluminum (panel b) again is limited by the [Al/Fe]=0.9$\pm$0.15 of our giant, and by [Al/Fe]=0 of the 1G, and the two purple lines pass through the extremes of the allowed $\alpha$\ range.
The range of allowed $\alpha$\ values imply a very specific range for the ejecta abundances, for helium 0.5$\lesssim Y_{max}\lesssim$1  and for aluminum 1.15$\lesssim$[Al/Fe]$_{max} \lesssim$1.55. 
The observed abundances should then be reproduced with  a single value of alpha. We also emphasize that similar specific constraints and correlations must exist for all the abundances of all other elements, such as Na and Mg, which are, unfortunately not available for the extreme star studied here. CCHP models should however address this issue and possibly be further tested by future observational determinations of other elements abundances for stars in the extreme second generation group.

We tentatively show implication of dilution by plotting the [Al/Fe]  and Y of the D group giants  at possible dilution values in the figure. We use for Y the values Y=0.32 and Y=0.30 consistent with the group D, from the interpretation of the middle main sequence. The plot shows that these few observations do not exclude the dilution model, although there are some difficulties in allocating the Y and [Al/Fe] abundances of the aluminum-rich D giants at the same dilution $\alpha$\ (compare the cyan dots in panels 2 and 3). 

While Fig.\,\ref{f3} deals with the requirements necessary to explain lithium in CCHP models, it also shows that  the lithium abundance requirements in the polluting gas are not very strict in the case of the giants in group D, as the ejecta in the forming gas are well diluted with pristine gas. At $\alpha >$0.4, it is difficult to distinguish between dilution with gas deprived from lithium or dilution with gas including lithium, if the measured abundance is $\sim 2$. For this ambiguity, it is not useful to compare the lithium abundances in the giants belonging to group D, in \cite{dorazi2015} sample, with the yields resulting in the AGB models, while we considered these stars for the comparison with the CCHP models.

\section{Discussion and conclusions}
\label{conclusions}

In the framework of the chromosome maps \citep{milone2017chromo} description of multiple populations in GCs, in this work we showed that a giant having quasi--standard lithium abundance \citep{dorazi2015} belongs to the group E of extreme stars in NGC\,2808, those having the largest possible helium abundance of 2G stars and the largest chemical anomalies signature of high temperature proton capture processing in the forming gas. Other stars with similar lithium belong to the less extreme group D in \cite{milone2015}. We attribute the giant \#46518 to the E group thanks to its unambiguous location in the pseudo--color (chromosome) maps deviced by \cite{milone2017chromo}, because spectroscopic data are available only for lithium and aluminium, and the aluminium abundance does not allow to easily discriminate between groups D and E, as discussed in \S\,\ref{2808li}.

In the AGB scenario, we identify the abundances of the E group with undiluted abundances of the ejecta of the most massive AGB and super--AGB stars \citep{dercole2008}, so, if  the giant \#46518 is made only by stellar ejecta, it is strictly necessary to have lithium synthetized and ejected into the intralcuster medium, together with the gas very highly processed by p-captures. On the contrary, it is not necessary to explain at the same time the D and C lithium abundances, as these groups are formed from ejecta of different mass progenitors \citep{dantona2016} diluted with pristine gas (see Fig.\,\ref{f3}, panel a). No strict correlation is required between the lithium abundance and the dilution. 

To understand the giant  \#46518 in the context of the AGB scenario, we computed super--AGB models evolutions.
Table\,\ref{tabyield} shows that the average lithium abundances in the ejecta of the super--AGB are compatible both with the abundances of p-capture elements (Mg, Al, Si) in the E population, and with the abundance $\log \epsilon$(Li)$\sim$1 in the giant \#46518. In particular, it is well reproduced by the 6.5\Msun\ evolution for $\eta$=0.01--0.02, but we have shown that the detailed lithium nucleosynthesis depends on the model detailed behaviour, so the agreement should be regarded as qualitative. We may expect variations of lithium,  as well as of other elements among the E group stars, if star formation did not occur in a unique burst after accumulation in the core cooling flow of all the remnants of the super--AGB ejecta of the different masses, but in quiet star formation events as suggested by the \cite{dercole2008} model.  The large spread in the star-to-star abundances of p--capture elements found among group E stars by \cite{carretta2018} is in fact consistent with the different abundances in Table\,\ref{tabyield} obtained from different mass progenitors, even for a single assumption on the mass loss law.
We further may expect lower lithium abundances in stars which may have suffered some extra--mixing below the bump level \citep{dicrisci2018}.

Different scenarios (the CCHP models) fully destroy lithium, so they are compatible with lithium abundances only for 2G stars whose signatures indicate significant ($>$40\%) dilution with pristine matter. 
We remark again that \cite{carretta2018} clearly show that a single dilution curve does not explain the whole observation set for the multiple populations of NGC\,2808. Actually, they propose that the extreme population is a result of CCHP massive stellar ejecta, while the rest of 2G stars would born on a longer timescale in the AGB
scenario. On the contrary, we focused mainly on the E population and concluded that the presence of a Li--rich giant and of the other typical chemical anomalies are more easily understood in the AGB scenario too.

We have shown in previous work, at several different levels, that the AGB scenario is capable of dealing with the abundances displayed in the globular cluster NGC\,2808 \citep{dercole2008, dercole2010, dantona2016, dicrisci2018}, and the abundance patterns of this cluster remain a Rosetta stone which must be dealt with to falsify the models for the formation of multiple populations. 
Our study shows that a `Lithium test' (even at a qualitative level) can provide key insight on the origin of multiple populations  and should be always considered when discussing new scenarios for the formation of multiple populations in GCs.

\section*{Acknowledgements }
This work has received funding from the European Research Council (ERC) under the European Union's Horizon 2020 research innovation programme (Grant Agreement ERC-StG 2016, No 716082 `GALFOR', PI: Milone), and the European Union's Horizon 2020 research and innovation programme under the Marie Sklodowska-Curie (Grant Agreement No 797100, beneficiary: Marino). APM and MT acknowledge support from MIUR through the the FARE project R164RM93XW SEMPLICE.

\bibliographystyle{aasjournal}

\begin{thebibliography}{}
\expandafter\ifx\csname natexlab\endcsname\relax\def\natexlab#1{#1}\fi

\bibitem[{{Asplund} {et~al.}(2006){Asplund}, {Lambert}, {Nissen}, {Primas}, \&
  {Smith}}]{asplund2006}
{Asplund}, M., {Lambert}, D.~L., {Nissen}, P.~E., {Primas}, F., \& {Smith},
  V.~V. 2006, \apj, 644, 229

\bibitem[{{Bastian} {et~al.}(2013){Bastian}, {Lamers}, {de Mink}, {Longmore},
  {Goodwin}, \& {Gieles}}]{bastian2013}
{Bastian}, N., {Lamers}, H.~J.~G.~L.~M., {de Mink}, S.~E., {et~al.} 2013,
  \mnras, 436, 2398

\bibitem[{{Bloecker}(1995)}]{blocker1995}
{Bloecker}, T. 1995, \aap, 297, 727

\bibitem[{{Bragaglia} {et~al.}(2010){Bragaglia}, {Carretta}, {Gratton},
  {Lucatello}, {Milone}, {Piotto}, {D'Orazi}, {Cassisi}, {Sneden}, \&
  {Bedin}}]{bragaglia2010}
{Bragaglia}, A., {Carretta}, E., {Gratton}, R.~G., {et~al.} 2010, \apjl, 720,
  L41

\bibitem[{{Cameron} \& {Fowler}(1971)}]{cameronfowler1971}
{Cameron}, A.~G.~W., \& {Fowler}, W.~A. 1971, \apj, 164, 111

\bibitem[{{Carretta}(2015)}]{carretta2015}
{Carretta}, E. 2015, \apj, 810, 148

\bibitem[{{Carretta} {et~al.}(2018){Carretta}, {Bragaglia}, {Lucatello},
  {Gratton}, {D'Orazi}, \& {Sollima}}]{carretta2018}
{Carretta}, E., {Bragaglia}, A., {Lucatello}, S., {et~al.} 2018, \aap, 615, A17

\bibitem[{{Carretta} {et~al.}(2009){Carretta}, {Bragaglia}, {Gratton},
  {Lucatello}, {Catanzaro}, {Leone}, {Bellazzini}, {Claudi}, {D'Orazi},
  {Momany}, {Ortolani}, {Pancino}, {Piotto}, {Recio-Blanco}, \&
  {Sabbi}}]{carretta2009a}
{Carretta}, E., {Bragaglia}, A., {Gratton}, R.~G., {et~al.} 2009, \aap, 505,
  117

\bibitem[{{Chantereau} {et~al.}(2016){Chantereau}, {Charbonnel}, \&
  {Meynet}}]{chantereau2016}
{Chantereau}, W., {Charbonnel}, C., \& {Meynet}, G. 2016, \aap, 592, A111

\bibitem[{{Charbonnel} \& {Zahn}(2007)}]{charbonnelzahn2007}
{Charbonnel}, C., \& {Zahn}, J.-P. 2007, \aap, 467, L15

\bibitem[{{Choi} {et~al.}(2016){Choi}, {Dotter}, {Conroy}, {Cantiello},
  {Paxton}, \& {Johnson}}]{choi2016}
{Choi}, J., {Dotter}, A., {Conroy}, C., {et~al.} 2016, \apj, 823, 102

\bibitem[{{D'Antona} {et~al.}(2005){D'Antona}, {Bellazzini}, {Caloi}, {Pecci},
  {Galleti}, \& {Rood}}]{dantona2005}
{D'Antona}, F., {Bellazzini}, M., {Caloi}, V., {et~al.} 2005, \apj, 631, 868

\bibitem[{{D'Antona} {et~al.}(2012){D'Antona}, {D'Ercole}, {Carini},
  {Vesperini}, \& {Ventura}}]{dantona2012li}
{D'Antona}, F., {D'Ercole}, A., {Carini}, R., {Vesperini}, E., \& {Ventura}, P.
  2012, \mnras, 426, 1710

\bibitem[{{D'Antona} \& {Ventura}(2007)}]{dantonaventura2007}
{D'Antona}, F., \& {Ventura}, P. 2007, \mnras, 379, 1431

\bibitem[{{D'Antona} {et~al.}(2016){D'Antona}, {Vesperini}, {D'Ercole},
  {Ventura}, {Milone}, {Marino}, \& {Tailo}}]{dantona2016}
{D'Antona}, F., {Vesperini}, E., {D'Ercole}, A., {et~al.} 2016, \mnras, 458,
  2122

\bibitem[{{D'Cruz} {et~al.}(1996){D'Cruz}, {Dorman}, {Rood}, \&
  {O'Connell}}]{dcruz1996}
{D'Cruz}, N.~L., {Dorman}, B., {Rood}, R.~T., \& {O'Connell}, R.~W. 1996, \apj,
  466, 359

\bibitem[{{de Mink} {et~al.}(2009){de Mink}, {Pols}, {Langer}, \&
  {Izzard}}]{demink2009}
{de Mink}, S.~E., {Pols}, O.~R., {Langer}, N., \& {Izzard}, R.~G. 2009, \aap,
  507, L1

\bibitem[{{Decressin} {et~al.}(2007){Decressin}, {Meynet}, {Charbonnel},
  {Prantzos}, \& {Ekstr{\"o}m}}]{decressin2007}
{Decressin}, T., {Meynet}, G., {Charbonnel}, C., {Prantzos}, N., \&
  {Ekstr{\"o}m}, S. 2007, \aap, 464, 1029

\bibitem[{{Denissenkov} \& {Hartwick}(2014)}]{denissenkov2014}
{Denissenkov}, P.~A., \& {Hartwick}, F.~D.~A. 2014, \mnras, 437, L21

\bibitem[{{D'Ercole} {et~al.}(2012){D'Ercole}, {D'Antona}, {Carini},
  {Vesperini}, \& {Ventura}}]{dercole2012}
{D'Ercole}, A., {D'Antona}, F., {Carini}, R., {Vesperini}, E., \& {Ventura}, P.
  2012, \mnras, 423, 1521

\bibitem[{{D'Ercole} {et~al.}(2010){D'Ercole}, {D'Antona}, {Ventura},
  {Vesperini}, \& {McMillan}}]{dercole2010}
{D'Ercole}, A., {D'Antona}, F., {Ventura}, P., {Vesperini}, E., \& {McMillan},
  S.~L.~W. 2010, \mnras, 407, 854

\bibitem[{{D'Ercole} {et~al.}(2008){D'Ercole}, {Vesperini}, {D'Antona},
  {McMillan}, \& {Recchi}}]{dercole2008}
{D'Ercole}, A., {Vesperini}, E., {D'Antona}, F., {McMillan}, S.~L.~W., \&
  {Recchi}, S. 2008, \mnras, 391, 825

\bibitem[{{Di Criscienzo} {et~al.}(2018){Di Criscienzo}, {Ventura}, {D'Antona},
  {Dell'Agli}, \& {Tailo}}]{dicrisci2018}
{Di Criscienzo}, M., {Ventura}, P., {D'Antona}, F., {Dell'Agli}, F., \&
  {Tailo}, M. 2018, \mnras, 479, 5325

\bibitem[{{Doherty} {et~al.}(2014){Doherty}, {Gil-Pons}, {Lau}, {Lattanzio},
  {Siess}, \& {Campbell}}]{doherty2014}
{Doherty}, C.~L., {Gil-Pons}, P., {Lau}, H.~H.~B., {et~al.} 2014, \mnras, 441,
  582

\bibitem[{{D'Orazi} {et~al.}(2015){D'Orazi}, {Gratton}, {Angelou}, {Bragaglia},
  {Carretta}, {Lattanzio}, {Lucatello}, {Momany}, {Sollima}, \&
  {Beccari}}]{dorazi2015}
{D'Orazi}, V., {Gratton}, R.~G., {Angelou}, G.~C., {et~al.} 2015, \mnras, 449,
  4038

\bibitem[{{Georgy} {et~al.}(2013){Georgy}, {Ekstr{\"o}m}, {Granada}, {Meynet},
  {Mowlavi}, {Eggenberger}, \& {Maeder}}]{georgy2013}
{Georgy}, C., {Ekstr{\"o}m}, S., {Granada}, A., {et~al.} 2013, \aap, 553, A24

\bibitem[{{Gieles} {et~al.}(2018){Gieles}, {Charbonnel}, {Krause},
  {H{\'e}nault-Brunet}, {Agertz}, {Lamers}, {Bastian}, {Gualandris}, {Zocchi},
  \& {Petts}}]{gieles2018}
{Gieles}, M., {Charbonnel}, C., {Krause}, M.~G.~H., {et~al.} 2018, \mnras, 478,
  2461

\bibitem[{{Iben}(1964)}]{iben1964firstdu}
{Iben}, Jr., I. 1964, \apj, 140, 1631

\bibitem[{{Milone} {et~al.}(2015){Milone}, {Marino}, {Piotto}, {Renzini},
  {Bedin}, {Anderson}, {Cassisi}, {D'Antona}, {Bellini}, {Jerjen},
  {Pietrinferni}, \& {Ventura}}]{milone2015}
{Milone}, A.~P., {Marino}, A.~F., {Piotto}, G., {et~al.} 2015, \apj, 808, 51

\bibitem[{{Milone} {et~al.}(2017){Milone}, {Piotto}, {Renzini}, {Marino},
  {Bedin}, {Vesperini}, {D'Antona}, {Nardiello}, {Anderson}, {King}, {Yong},
  {Bellini}, {Aparicio}, {Barbuy}, {Brown}, {Cassisi}, {Ortolani}, {Salaris},
  {Sarajedini}, \& {van der Marel}}]{milone2017chromo}
{Milone}, A.~P., {Piotto}, G., {Renzini}, A., {et~al.} 2017, \mnras, 464, 3636

\bibitem[{{Piotto} {et~al.}(2007){Piotto}, {Bedin}, {Anderson}, {King},
  {Cassisi}, {Milone}, {Villanova}, {Pietrinferni}, \& {Renzini}}]{piotto2007}
{Piotto}, G., {Bedin}, L.~R., {Anderson}, J., {et~al.} 2007, \apjl, 661, L53

\bibitem[{{Renzini} {et~al.}(2015){Renzini}, {D'Antona}, {Cassisi}, {King},
  {Milone}, {Ventura}, {Anderson}, {Bedin}, {Bellini}, {Brown}, {Piotto}, {van
  der Marel}, {Barbuy}, {Dalessandro}, {Hidalgo}, {Marino}, {Ortolani},
  {Salaris}, \& {Sarajedini}}]{renzini2015}
{Renzini}, A., {D'Antona}, F., {Cassisi}, S., {et~al.} 2015, \mnras, 454, 4197

\bibitem[{{Scalo} \& {Ulrich}(1975)}]{scalo1975}
{Scalo}, J.~M., \& {Ulrich}, R.~K. 1975, \apj, 200, 682

\bibitem[{{Siess}(2010)}]{siess2010}
{Siess}, L. 2010, \aap, 512, A10

\bibitem[{{Ventura}(2010)}]{ventura2010}
{Ventura}, P. 2010, in IAU Symposium, Vol. 268, Light Elements in the Universe,
  ed. C.~{Charbonnel}, M.~{Tosi}, F.~{Primas}, \& C.~{Chiappini}, 147--152

\bibitem[{{Ventura} \& {D'Antona}(2005)}]{ventura2005a}
{Ventura}, P., \& {D'Antona}, F. 2005, \aap, 431, 279

\bibitem[{{Ventura} \& {D'Antona}(2008)}]{ventura2008aa}
---. 2008, \aap, 479, 805

\bibitem[{{Ventura} \& {D'Antona}(2011)}]{ventura2011sagb}
---. 2011, \mnras, 410, 2760

\bibitem[{{Ventura} {et~al.}(2000){Ventura}, {D'Antona}, \&
  {Mazzitelli}}]{ventura2000}
{Ventura}, P., {D'Antona}, F., \& {Mazzitelli}, I. 2000, \aap, 363, 605

\bibitem[{{Ventura} {et~al.}(2001){Ventura}, {D'Antona}, {Mazzitelli}, \&
  {Gratton}}]{ventura2001}
{Ventura}, P., {D'Antona}, F., {Mazzitelli}, I., \& {Gratton}, R. 2001, \apjl,
  550, L65

\end{thebibliography}

\end{document}